\documentclass[lettersize,journal]{IEEEtran}
\usepackage{amsmath,amsfonts}
\usepackage{algorithmic}
\usepackage{algorithm}
\usepackage{array}
\usepackage[caption=false,font=normalsize,labelfont=sf,textfont=sf]{subfig}
\usepackage{textcomp}
\usepackage{stfloats}
\usepackage{url}
\usepackage{ulem}
\usepackage{verbatim}
\usepackage{graphicx}
\usepackage{cite}
\begin{document}

\title{GainNet: Coordinates the Odd Couple of Generative AI and 6G Networks}
%
\author{Ning~Chen,~\IEEEmembership{}
       Jie~Yang,~\IEEEmembership{}
       Zhipeng~Cheng,~\IEEEmembership{}
       Xuwei~Fan,~\IEEEmembership{}
       Zhang~Liu,~\IEEEmembership{}
       Bangzhen~Huang,~\IEEEmembership{}
       Yifeng~Zhao,~\IEEEmembership{}
       Lianfen~Huang,~\IEEEmembership{}\\
        Xiaojiang~Du,~\IEEEmembership{Fellow,~IEEE,}
        and~Mohsen~Guizani,~\IEEEmembership{Fellow,~IEEE}
\thanks{Corresponding author: Yifeng Zhao (email:zhaoyf@xmu.edu.cn)}
\thanks{Ning Chen, (email: ningchen@stu.xmu.edu.cn), Xuwei Fan (email: xwfan@stu.xmu.edu.cn), Zhang Liu (email: zhangliu@stu.xmu.edu.cn), Bangzhen Huang (email: huangbz0714@stu.xmu.edu.cn), Yifeng Zhao (email:zhaoyf@xmu.edu.cn), and Lianfen Huang (email: lfhuang@xmu.edu.cn) are with the School of Informatics, Xiamen University, 361005 Xiamen, China.}
\thanks{Jie Yang (email: leoyang@stu.xmu.edu.cn) is with the National Institute for Data Science in Health and Medicine, Xiamen University, 361102 Xiamen, China.}
\thanks{Zhipeng Cheng (email: chengzp\_x@163.com) is with the School of Future Science and Engineering, Soochow University, 215006 Suzhou, China.}
\thanks{Xiaojiang Du (email: dxj@ieee.org) is with the Department of Electrical and Computer Engineering, Stevens Institute of Technology, Hoboken, NJ, USA.}
\thanks{Mohsen Guizani (email: mguizani@ieee.org) is with Mohamed Bin Zayed University of Artificial Intelligence, Abu Dhabi, UAE.}
}



\maketitle

\begin{abstract}
The rapid expansion of AI-generated content (AIGC) reflects the iteration from assistive AI towards generative AI (GAI) with creativity. Meanwhile, the 6G networks will also evolve from the Internet-of-everything to the Internet-of-intelligence with hybrid heterogeneous network architectures. In the future, the interplay between GAI and the 6G will lead to new opportunities, where GAI can learn the knowledge of personalized data from the massive connected 6G end devices, while GAI's powerful generation ability can provide advanced network solutions for 6G network and provide 6G end devices with various AIGC services. However, they seem to be an odd couple, due to the contradiction of data and resources. To achieve a better-coordinated interplay between GAI and 6G, the \uline{GAI}-\uline{n}ative \uline{net}works (GainNet), a GAI-oriented collaborative cloud-edge-end intelligence framework, is proposed in this paper. By deeply integrating GAI with 6G network design, GainNet realizes the positive closed-loop knowledge flow and sustainable-evolution GAI model optimization. On this basis, the \uline{GAI}-oriented generic \uline{r}esource \uline{o}rchestration \uline{m}echanism with \uline{i}ntegrated \uline{s}ensing, \uline{c}ommunication, and \uline{c}omputing (GaiRom-ISCC) is proposed to guarantee the efficient operation of GainNet. Two simple case studies demonstrate the effectiveness and robustness of the proposed schemes. Finally, we envision the key challenges and future directions concerning the interplay between GAI models and 6G networks.
\end{abstract}


\section{Introduction}
\IEEEPARstart{A}{I-generated} content (AIGC) refers to taking advantage of AI to achieve human-like automated content generation \cite{ref10}. As the key enabling technology of AIGC, generative AI (GAI) has been rapidly developed in recent years \cite{ref8}, emerging various underlying backbone architectures, e.g., generative adversarial networks (GAN), variational auto-encoders (VAE), diffusion model (DM) and Transformer \cite{ref4,ref5}. Based on this, technology giants have released a variety of foundation models for the format-various content processing and generation, such as OpenAI's GPT and DALL-E series, Meta's LLaMA, Google's BERT, etc\cite{ref12}. GAI model is characterized by the huge parameter scale, extensive knowledge extraction from the pre-training upon massive unlabeled data sets, and domain-across robustness with few-shot fine-tuning \cite{ref5,ref12}. However, despite GAI having gained tremendous development, it still faces the challenge of limited availability of high-quality public data \cite{ref2}.

%

On the other hand, the massively connected terminals in the B5G and 6G era, such as Internet of Things (IoT) devices, will generate explosive-growth distributed data. It is estimated that 30.9 billion IoT devices will generate 79.4 Zettabytes (ZB) of data \cite{ref15}. Meanwhile, 6G will adopt new enabling technologies and network architectures, e.g., the cell-free network architecture that can provide wide-area seamless connection and collaborative cloud-edge-end computing network for ubiquitous computing \cite{ref13}. Thus, benefiting from the growth of terminal data and the evolution of network architectures, in the coming B5G and 6G era, we will witness a shift in the networking paradigm from the Internet-of-everything to the Internet-of-intelligence, where native AI will sink from distant cloud servers to the edge of the network closer to users to form edge intelligence (EI) \cite{ref14}. In particular, to provide inclusive AI services, the 6G network can no longer simply consider data transmission, but is built as a comprehensive service platform with integrated sensing, communication, and Computing (ISCC) \cite{ref14}.

Thus, when GAI encounters 6G networks, the interplay between them will spawn new growth potentials. The massive data of IoT devices in the 6G network can alleviate the public data shortage of GAI, in turn, GAI can endow the 6G network with expert-free efficient network management, and enable the end devices to enjoy powerful context-aware content generation services \cite{ref4,ref5}. Whereas, as shown in TABLE \ref{table_1}, the combination of GAI and 6G faces two key challenges due to the discrepant characteristics. 1) Data sharing is blocked due to privacy concerns. Traditional GAI models are usually pre-trained and fine-tuned based on the collected massive data sets. However, to avoid privacy leakage, 6G users are unwilling to share their local data. 2) Terminal deployment is hindered by the huge model scale. GAI models with billions of parameters need to be pre-trained on large-scale datasets with huge amounts of computing and time resources \cite{ref5,ref9,ref10}, which is unaffordable for resource-constrained 6G end devices.

Facing the above challenges, federated split learning (FSL) emerges as an effective solution, which keeps the 6G end devices’ data locally and performs model splitting. Federated learning (FL) is a collaborative learning paradigm that can utilize the knowledge of distributed data under the premise of protecting users’ privacy \cite{ref2,ref11,ref14}. FSL can further decompose the GAI model, only deploy the lightweight part in the terminal, and keep the part with high computing power demand in the cloud with sufficient computing resources \cite{ref9,ref11}. In addition, federated prompt learning (FPL), which locks the GAI models’ backbone, and utilizes the distributed training of additional modules to guide GAI models, is also an effective strategy to acquire knowledge of end devices’ data. 

Even so, due to the domain-specific nature of the GAI model, it needs to perform fine-tuning in the direction specific to the desired application, for example, even if ChatGPT is used as the foundation model, the traffic domain concerns road safety information more, while the medical domain maybe more concerned with the suitability of the medicine for the condition. Meanwhile, the traditional GAI models experience a one-way development from top to bottom and from development to deployment, and the models' growth in the terminal cannot be fed back to the original model. Therefore, it is necessary to further find methods that can not only make use of the private data but also achieve more sustainable optimization and application of GAI models. Moreover, the optimization of sensing, communication, and computing is performed independently in a task-agnostic manner in traditional wireless networks. However, when facing GAI services, sensing, communication, and computing are deeply coupled, and the performance of any sub-process among them does not represent the final GAI’s quality of service (QoS) \cite{ref14}. Thus, it needs a higher-level evaluation metric that can represent the QoS of the GAI model and resource scheduling strategy following the metric. Therefore, based on the above analysis, this paper focuses on the following two key issues in the interplay between GAI and 6G.

\textit{1) How to build a sustainable knowledge extraction pipeline between the GAI model and the 6G terminals? }

\textit{2) How to reasonably schedule resources to guarantee the QoS of the GAI model?}

\begin{table*}[htbp]
\center
\footnotesize
\caption{Comparison of differential features between GAI and 6G networks.}
\label{table_1}
\renewcommand\arraystretch{1.5}
\begin{tabular}{|c|c|c|c|c|}
\hline
& \textbf{Data} & \textbf{Resource} & \textbf{Construction} & \textbf{Dynamicity} \\ \hline
\textbf{GAI} & Massive / Balanced / Public / Stale & Sufficient / Consistent & Isomorphic / Centralized & Low \\ \hline
\textbf{6G} & Slight / Unbalanced / Personalized / Fresh & Limited / Uneven & Heterogeneous / Distributed & High \\ \hline
\end{tabular}
\end{table*}

To solve the above problems, firstly, we propose the \uline{GAI}-\uline{n}ative \uline{net}works (GainNet), which constructs the knowledge transmission pipeline between cloud GAI models and end lightweight models. Furthermore, the \uline{GAI}-oriented generic \uline{r}esource \uline{o}rchestration \uline{m}echanism with \uline{i}ntegrated \uline{s}ensing, \uline{c}ommunication, and \uline{c}omputing (GaiRom-ISCC) is proposed. To the best of our knowledge, this paper is among the first to study the coordination of GAI and 6G by collaborative cloud-edge-end intelligence with the resource orchestration that creates direct resource-model mapping. Major contributions are summarized as follows:

$\bullet$ GainNet, a collaborative cloud-edge-end intelligence framework, is proposed, which enables an integrated design of GAI and 6G networks, realizing positive closed-loop knowledge flow and sustainable-evolution GAI model optimization.

$\bullet$ To guarantee the efficient resource provision of the proposed GainNet framework, we propose the GaiRom-ISCC, which accomplishes user association and resource allocation with one stone.

$\bullet$ Simulation results for two case studies demonstrate the effectiveness and robustness of the proposed schemes.

$\bullet$ We discuss the key challenges and future directions for the interplay between GAI and 6G networks.


\section{Sustainable-evolution Collaborative GAI-native Networks}
In this section, we propose the GainNet, a collaborative cloud-edge-end intelligence network framework for GAI models deployed in the 6G networks, which can realize positive closed-loop knowledge flow and sustainable-evolution GAI model optimization.

\subsection{Overview of GAI Models}
The success of GAI relies on the innovation of backbone model architectures, such as the famous Transformer in the field of natural language processing (NLP) proposed by Google \cite{ref1}, which is the backbone of many state-of-the-art foundation models, e.g., GPT-3, DALL-E 2, Codex, Gopher, etc. Compared with the traditional RNN-based model, the Transformer based on the self-attention mechanism has better parallelism and fewer data requirements \cite{ref1}. Typically, the lifecycle of a GAI model contains pre-training, fine-tuning, and inference. First, the pre-training on large-scale unlabeled data sets endows the GAI models with strong domain-across comprehensive abilities, nevertheless, the pre-training of GAI models with billions of parameters requires huge computing and time resources \cite{ref9}, which is unaffordable for resource-limited 6G end devices \cite{ref5}. Then, the fine-tuning for different vertical domains enables the original GAI model to further increase its domain-specific knowledge. Conventional centralized fine-tuning relies on continuous data collection from users, which violates personal data privacy \cite{ref5}. Last, the inference is more inclined to the application of the GAI model, and output text and images that meet users' intentions. However, GAI-based inference is affected by the domain relevance and the uncertainty of the network environment \cite{ref5}.

\subsection{Collaborative Cloud-edge-end Intelligence}
Collaborative cloud-edge-end intelligence for the training and inference service of AI models has attracted a lot of attention \cite{ref6,ref15}, in particular, where edge intelligence (EI) can enable the sinking of AI from the central cloud to the edge of the network closer to users’ devices \cite{ref14,ref15,ref9,ref11}.

The advent of the computing power network and native AI framework in 6G networks establishes the feasibility of network architecture and computing power resources for deploying GAI models in 6G networks. However, compared with traditional AI models, GAI exhibits new characteristics when it encounters 6G, including  \textit{1) GAI model needs to be pre-trained based on the large-size datasets}, \textit{2) domain-specific fine-tuning can unlock greater performance potential of GAI}, and \textit{3) due to the large size of the GAI model, it cannot be fully deployed on resource-constrained 6G end devices}.

\subsection{Sustainable-evolution GainNet}

\begin{figure*}[!t]
\centering
\includegraphics[width=6.0in]{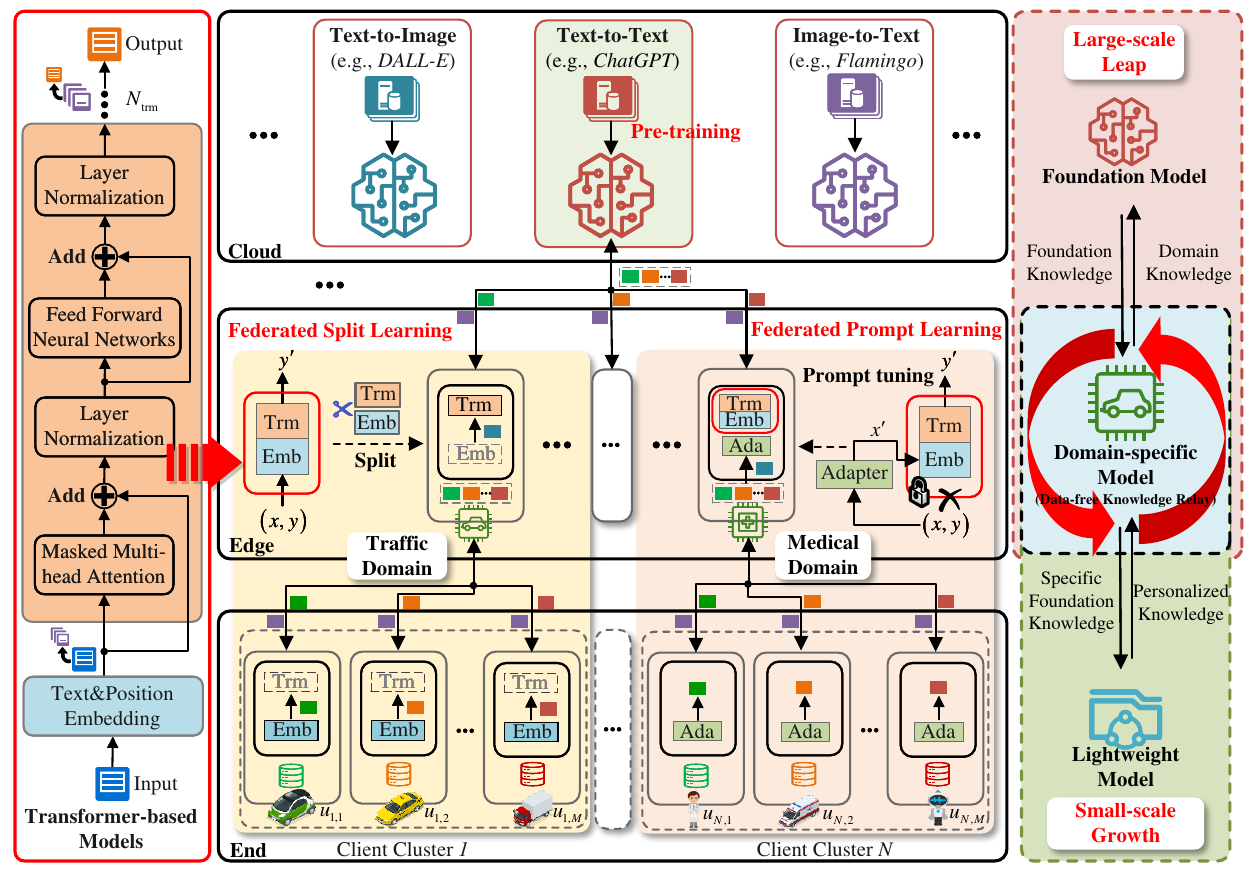}
\caption{The architecture of the proposed GainNet. Since the GAI model cannot be fully deployed on 6G terminal devices, to obtain personalized knowledge of local data of 6G terminals, the FL framework of distributed EI is adopted, and there are two solutions. The first is the direct fine-tuning based on FSL, which performs split learning on the GAI model, keeps the modules with high computing power demand (e.g. the Transformer layer of ChatGPT) on the edge server, and puts the lightweight modules with low computing power demand (e.g. the Embedding layer) on the 6G end devices. The other is FPL to perform indirect fine-tuning, that is, locking the original GAI model and adding additional modules to guide the optimization of the GAI models, such as adding external Adapters.}
\label{fig_1}
\end{figure*}

Fig. \ref{fig_1} shows the proposed GainNet framework. GainNet is a sustainably evolving collaborative cloud-edge-end intelligence network tailored for GAI in 6G networks. First, the \textit{foundation models} of GAI are deployed on the cloud servers, which provide large-scale data sets and massive computing resources for pre-training. Then, the \textit{domain-specific models} are deployed on edge servers close to the terminals, and in league with 6G end devices’ distributed local \textit{lightweight models}, fine-tuning is performed upon the data of multiple applications, achieving better adapt to downstream tasks. With the federated framework, 6G end devices realize the extraction of local data knowledge without data sharing with privacy and security threats. 

%
%

Specifically, in the above process of cloud-edge-end cooperation to perform GAI model optimization, the edge server acts as the data-free knowledge relay to build a bridge between the cloud server and the terminals and further divides the GainNet’s sub-layers into functional logical sub-networks for GAI optimization, i.e. cloud-edge subnetwork and edge-end subnetwork. The tasks assignment of the two logical subnetworks are as follows.

$\bullet$ \textbf{Domain-across large-scale leaps in cloud-edge subnetworks.} The cloud server utilizes the local large-size datasets and massive computing resources for pre-training to endow GAI models with powerful foundation knowledge, which will be copied to the cluster of edge servers in the process of delivering the model. Meanwhile, the edge servers gather the personalized knowledge learned from the distributed terminals to obtain new domain knowledge and feedback to the cloud server to further enhance the performance of the GAI model in the cloud.

$\bullet$ \textbf{Domain-specific small-scale growth in edge-end subnetworks.} The edge server associates the terminal cluster with the cell-free structure and provides professional knowledge of the domain-specific model to terminals, which is obtained from cloud services and undergoing fine-tuning for a specific domain. Meanwhile, the personalized knowledge in 6G end devices’ local data is collected and summarized by the edge server to obtain new domain knowledge.

Utilizing the collaboration between the GAI foundation models deployed in the cloud servers, the domain-specific models in the edge servers, and the local lightweight models in the 6G end devices, positive closed-loop knowledge flow and sustainable-evolution GAI model optimization can be realized. Concretely, as a bidirectional data-free knowledge relay, the edge server inherits the large-scale enhanced popular knowledge obtained from the large-size datasets in the cloud and provides high-quality and low-cost domain-specific inference services for 6G end devices. Furthermore, the personalized knowledge from the local data of the terminal devices can be transferred upward by the edge server, and the comprehensive performance of the cloud GAI model can be enhanced at a small scale. GainNet realizes the closed-loop bidirectional flow of knowledge between the cloud and the terminals and creates recurrent positive feedback for GAI model optimization. Thus, GainNet can enable \textit{sustainable and efficient evolution (pre-training and fine-tuning)} and \textit{high-quality and low-cost application (inference)} of GAI on 6G networks.

\section{GAI-oriented Generic Resource Orchestration With Integrated Sensing, Communication, and Computing}

In this section, to guarantee the efficient operation of the proposed GainNet, the GaiRom-ISCC is proposed, which accomplishes user association and resource allocation with one stone.

\subsection{Model Optimization with Integrated Sensing, Communication, and Computing}

\begin{figure*}[t]
\centering
\includegraphics[width=6.0in]{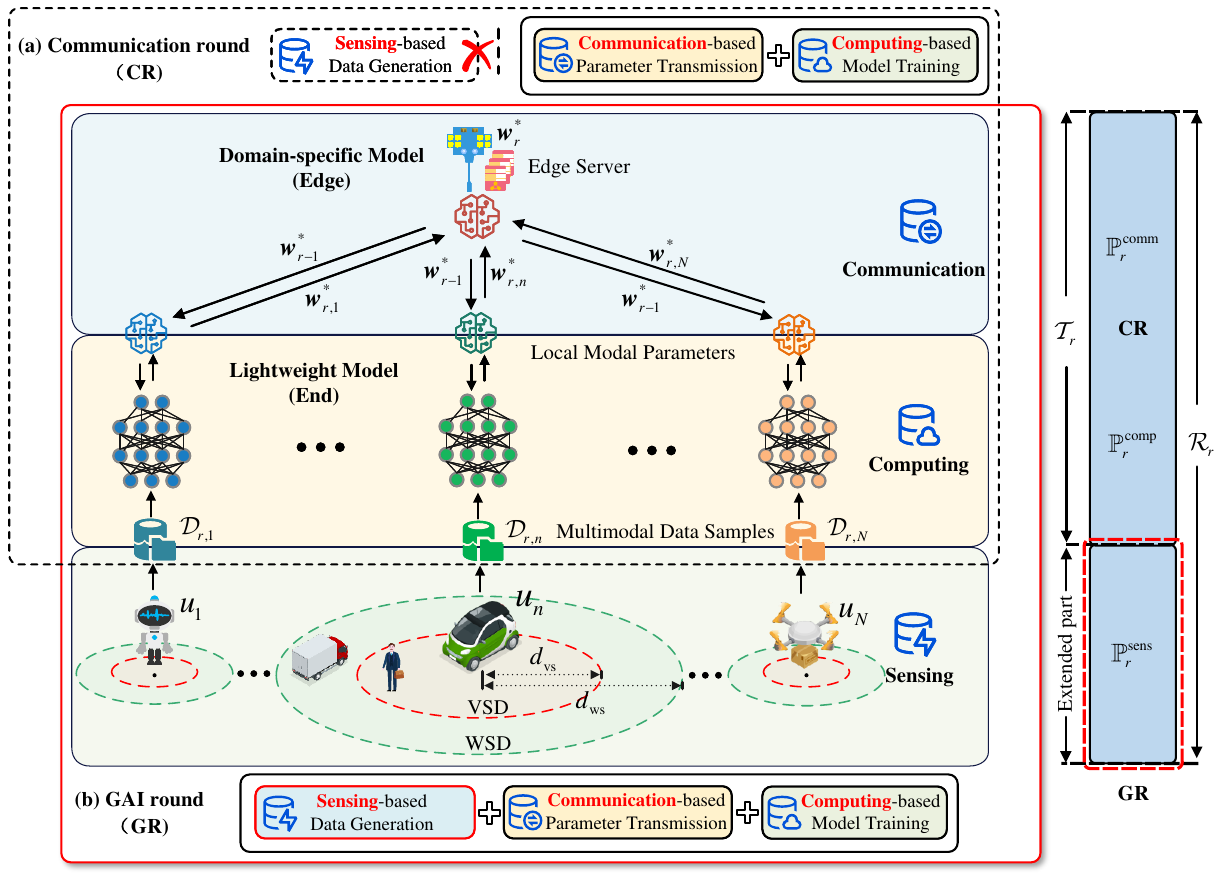}
\caption{Optimization of GAI edge-end subnetworks with integrated sensing, communication, and computing. Since the deployment locations of cloud servers and edge servers are relatively fixed and there are sufficient resources, the optimization of GainNet’s cloud-edge subnetworks is stable. Due to the limited resources, high mobility, and changes in data distribution of 6G terminals, the distributed optimization of edge-end subnetworks is full of uncertainty. Thus, this paper focuses on the optimization of the GAI model on edge-end subnetworks and the corresponding resource provisioning. In particular, the edge domain-specific model and end lightweight model are collectively referred to as the \textit{edge-end model}.}
\label{fig_2}
\end{figure*}

The independent optimization of sensing, communication, and computing in traditional wireless networks is not suitable for the GainNet deployed in 6G networks \cite{ref14}. On the one hand, the sensing, communication, and computing processes in GainNet are not entirely independent and parallel, instead, they involve complex coupling relationships. Firstly, there are complex game relationships caused by resource reuse, such as sensing, communication, and computing processes are inseparable from time resources, wireless sensing (WS) and wireless communication (WC) both need spectrum resources, etc. Secondly, there are compulsory serial timing constraints (CSTCs) caused by the sequential temporal relationship. For example, the uplink communication transmits the trained model output from the computing progress, and the computing needs to be based on the initial model transmitted by the downlink communication and the sensed data brought from sensing \cite{ref16}. On the other hand, the single process performance of sensing, communication, and computing cannot represent the QoS of GainNet independently. For example, spectral efficiency, which represents communication performance, is not a human-perceivable metric for GAI service.

Therefore, in this paper, we consider the QoS-oriented optimization of the GAI model with ISCC in 6G networks. Consider a GainNet edge-end subnetwork based on the client-server (CS) architecture, which consists of a client set $\mathcal{U}=\left\{ {{u}_{1}},{{u}_{2}},\ldots ,{{u}_{N}} \right\}$ composed of $N$ 6G end devices and an edge server acting as the model aggregator, as shown in Fig. \ref{fig_2}. Fully considering the differential resource consumption and the influence of personalized data on the GAI model, we incorporate the sensing process into the optimization of the GAI model based on the FL framework. The communication round (CR) in traditional FL is extended to the GAI round (GR) as shown in Fig. \ref{fig_2}. Therefore, for GR ${{\mathcal{R}}_{r}}\in \mathcal{R}$, GAI model optimization contains three sub-processes of sensing $\mathbb{P}_{r}^{\text{sens}}$, communication $\mathbb{P}_{r}^{\text{comm}}$, and computing $\mathbb{P}_{r}^{\text{comp}}$.

$\bullet$ \textbf{Sensing-based data generation.} Multimodal sensing data ${{\mathcal{D}}_{r,n}}$ that can be used for local model training and knowledge extraction is generated based on the sensing progress. For example, camera-based visual sensing (VS) and integrated sensing and communication (ISAC)-based wireless sensing (WS) with complementary advantages can be considered, and their effective data acquisition areas are defined as visual sensing domain (VSD) ${{S}_{\text{vs}}}$ and wireless sensing domain (WSD) ${{S}_{\text{ws}}}$ respectively \cite{ref14}.

$\bullet$ \textbf{Communication-based parameter transmission.} It can be divided into two sub-modules in opposite directions, i.e., edge model distribution $\vec{\mathbb{P}}_{r}^{\text{comm}}$ and local model aggregation $\overset{\scriptscriptstyle\leftarrow}{\mathbb{P}}_{r}^{\text{comm}}$, which complete the distribution of edge domain-specific knowledge and the upload of local personalized knowledge, respectively.

$\bullet$ \textbf{Computing-based model training.} 6G end devices that are selected as working clients utilize the data ${{\mathcal{D}}_{r,n}}$ to perform model training to minimize the local loss $\ell ({{w}_{r,n}})$. Finally, the optimal local model of each client performs parameter aggregation on the edge server to minimize the global loss function $\ell ({{w}_{r}})$ to obtain the optimal domain-specific model $w_{r}^{*}$ of GR ${{\mathcal{R}}_{r}}$.

\subsection{GaiRom-ISCC}

\begin{figure*}[t]
\centering
\includegraphics[width=6.0in]{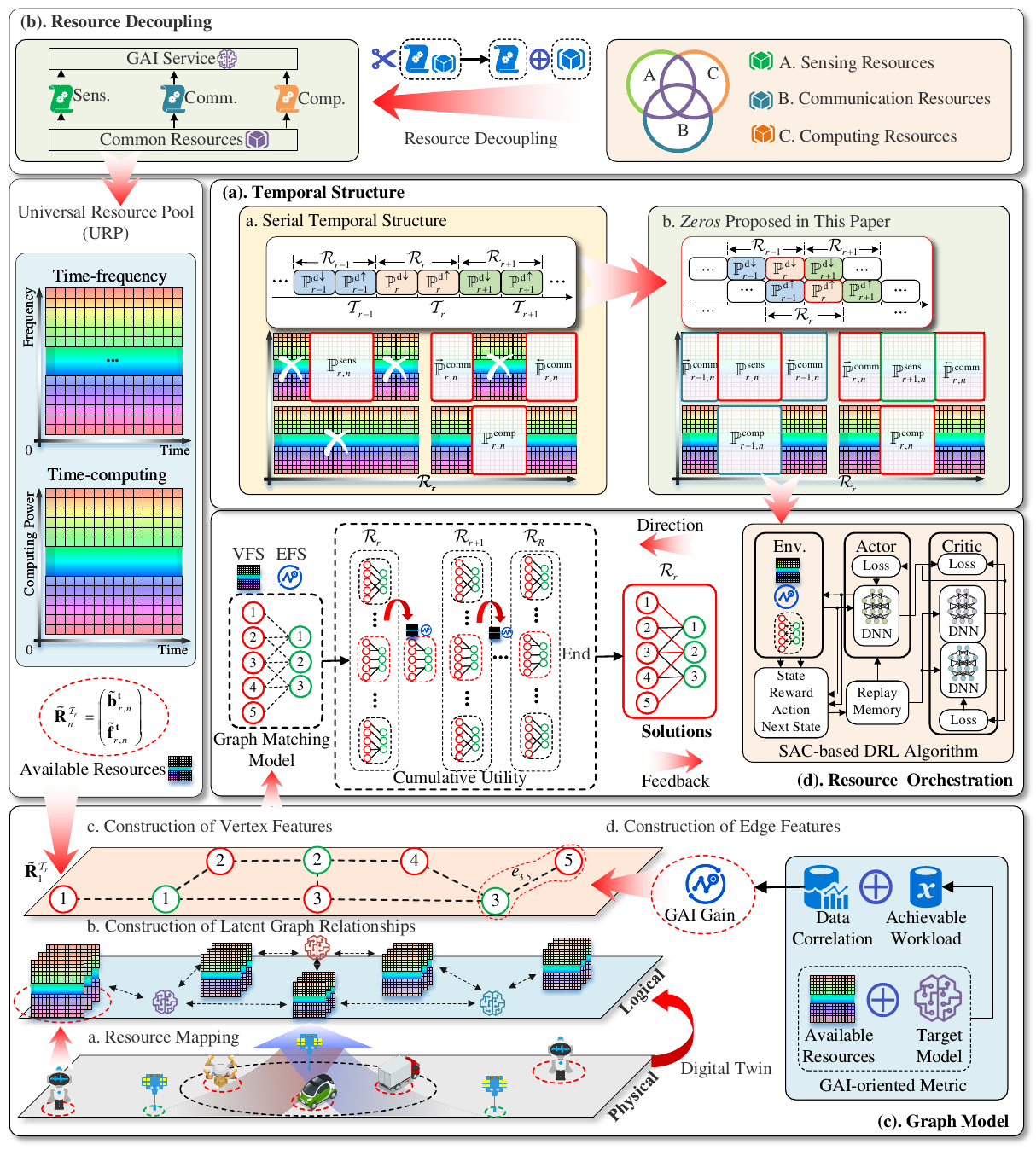}
\caption{Overview of the proposed GaiRom-ISCC. First, based on Zeros, a time-efficient temporal structure, partial parallel execution of sensing, communication, and computing between neighboring GRs can be achieved, which reduces the resource waste caused by the resource idle due to the CSTCs between the intra-GR sensing, communication, and computing and improves the resource utilization rate. Then, based on the decoupled URPs, the multi-domain resources can be invoked without distinction among sensing, communication, and computing, creating the resource objects for the mapping between resources and GAI models. Furthermore, the graph-based representation of the resource-model relationship is used to fully explore the relationship characteristics between the clients’ resources and GAI models. Finally, the GAI-oriented generic resource orchestration is implemented with DRL-based graph matching.}
\label{fig_3}
\end{figure*}


Resource management of the GainNet network mainly consists of two sub-problems, one is the association between domain-specific models and candidate 6G terminals, and the other is the internal resources, such as time resources, frequency resources, and computing power resources, are scheduled among the sensing, communication, and computing. The two problems above mentioned are coupled because the association results affect the internal allocation of local resources, and the methods of forcing decoupling can only obtain a suboptimal solution. Fortunately, from the perspective of resources, the above two sub-problems can be combined by completing the orchestration of local resources to domain-specific models deployed in edge servers \cite{ref6, ref10}. Therefore, we propose the GaiRom-ISCC, which directly executes the matching between the decoupled universal resource pool (URP) and the GAI edge models. The key components of the GaiRom-ISCC are as follows.

\subsubsection{Z-shaped overlapped temporal structure}

The three sub-processes of GAI model optimization for GR ${{\mathcal{R}}_{r}}\in \mathcal{R}$ can be summarized into two parts, i.e. \textit{data generation} $\mathbb{P}_{r}^{\text{d}\downarrow }$ and \textit{data consumption} $\mathbb{P}_{r}^{\text{d}\uparrow }$. Among them, extended sensing $\mathbb{P}_{r}^{\text{sens}}$ completes the task of data generation, while the communication $\mathbb{P}_{r}^{\text{comm}}$ and computing $\mathbb{P}_{r}^{\text{comp}}$ that constitutes the traditional CR are dedicated to model training, which can be understood as data consumption. From the perspective of sequential relationships, there are CSTCs between the data generation and data consumption of each GR, which bring great constraints and limits to the parallelism of model optimization and the flexibility of resource allocation. Fortunately, we find that there are constraints inside each GR, but the situation changes greatly when extending the view to the adjacent GR. Considering the complexity of data generation and the high data quality requirements of GAI model optimization, timing alignment of data generation in the post-GR and data consumption in the pre-GR is a reasonable and efficient choice.

Inspired by the discussion above, we propose the \uline{Z}-shap\uline{e}d ove\uline{r}lapped temp\uline{o}ral \uline{s}tructure (Zeros) for GAI model optimization, in which neighboring GRs are no longer executed in a complete serial pattern, but arranged in a partially overlapping Z-shaped pattern. The data generation $\mathbb{P}_{r}^{\text{d}\downarrow }$ of GR ${{\mathcal{R}}_{r}}$ and the data consumption $\mathbb{P}_{r-1}^{\text{d}\uparrow }$ of GR ${{\mathcal{R}}_{r-1}}$ implement a GR-across overlap similar to Tetris in the same CR.

\subsubsection{Decoupled Universal Resource Pool Model}

According to the above analysis, there is a partial overlap among the time, frequency, and computing power resources required by the sensing, communication, and computing involved in the GAI service, which brings great complexity and challenges to resource management. To this end, we propose the URP model that decouples multi-domain physical resources with sensing, communication, and computing. Specifically, each URP is constituted by a CR, which is shared by neighboring GRs. Meanwhile, it is not exclusive to sensing, communication, and computing, but they can all equally use the resources in the resource pool. Furthermore, since both frequency and computing power resources have time attributes, we integrate one-dimensional time resources into the above two resources to obtain a two-dimensional URP consisting of a two-dimensional time-frequency resource matrix $\mathbf{\tilde{b}}_{r,n}^{\mathbf{t}}$ and a two-dimensional time-computing resource matrix $\mathbf{\tilde{f}}_{r,n}^{\mathbf{t}}$.

\subsubsection{Graph-based Representation of the Resource-model Relationship}

After obtaining the URP, we can explore the relationship between the resource and the domain-specific models. Since the main goal of the 6G terminals is to enhance the performance of the GAI model, we define the \textit{learning performance gain} of GAI models as the evaluation metric representing QoS \cite{ref16}. The achievable contribution of a client to GAI depends on the achievable workload quantified by the number of data samples consumed by its local training, as well as the degree of correlation between the local lightweight model and the edge domain-specific model \cite{ref2}. Specifically, the clients’ workload depends on the resources in the URP because the workload is the final output after performing the sensing, communication, and computing. Meanwhile, since the communication sub-process is completed by the client and the edge server together, affected by the quality of the transmission path, the resource consumption of the communication between different end-edge pairs is different. Therefore, the workload is also affected by the edge server. Under the premise of determining the client-edge server pair, the workload achievable between the local URP and the edge server is a convex optimization problem with a mixture of inequality and equality constraints, and the optimal solution can be obtained by the Lagrange multiplier method \cite{ref16}. That is, the achievable workload of a client is jointly determined by its available resources (i.e., URP) and target model (i.e., different edge servers), and its achievable workload in a specific combination is determined. In addition, the degree between models can be expressed by the correlation of the data they were trained on, such as the variance or relative entropy of the data distribution. Therefore, GAI gain is jointly determined by data correlation and achievable workload. Based on the above definition, the graph-based representation of the resource-model relationship consists of the following procedures.

$\bullet$ \textbf{Resource mapping}. The physical resources of multiple domains are represented as two-dimensional URP composed of the two-dimensional time-frequency resource matrix and time-computing resource matrix.

$\bullet$ \textbf{Construction of latent relationships graph.} The latent relationships graph $\mathcal{G}=(\mathcal{V},\mathcal{E})$ is established between the 6G terminals and the edge models. The vertex set $\mathcal{V}$ consists of 6G end devices and domain-specific models deployed on edge servers, and the potential edge set $\mathcal{E}$ is represented by the connectable link between the 6G terminals and domain-specific models.

$\bullet$ \textbf{Representation of vertex features.} The URP is modeled as the unique feature of the vertex, in which the weight of vertices represents the available resources of the 6G terminals.

$\bullet$ \textbf{Representation of edge features.} The deterministic GAI learning performance gain between the 6G end devices and the edge models located at both ends of the potential edge, which represents the QoS of the GAI service, is taken as the weight of the edge, which is the feature of the edge.

Based on the above process, the latent graph relationship, vertex features, and edge features together can obtain the latent graph matching model, which contains the prior knowledge of vertex feature sets (VFS) and edge feature sets (EFS). VFS represents the available resources of the 6G terminals, while EFS represents the GAI gain that can be realized on the matching relationship, which can reflect the value that can be created by the resource-model association.

\subsubsection{DRL-based Graph Matching for Generic Resource Orchestration}

After the Graph-based representation of the resource-model relationship, the user association problem and the resource management problem within sensing, communication, and computing are modeled as the resource orchestration problem involving resource-model mapping, i.e., “\textit{Which domain-specific model should the resources map to?}”. If the traditional serial temporal structure is adopted, the 6G terminals can establish an association with the edge model that can obtain the maximum GAI gain to complete the resource orchestration. However, due to the adoption of Zeros, neighboring GRs are not independent, but there is a coupling relationship of resource sharing, that is, the GAI gain of neighboring GRs is correlated, and this problem is NP-hard and difficult to solve by using traditional optimization methods.

Deep reinforcement learning (DRL) is shown to be a competitive candidate for solving resource management problems \cite{ref4,ref10}. In particular, the resource orchestration problem of GainNet is adjacent-process-dependent, that is, the result of each GR is only affected by the previous GR and independent of the others. At the same time, different terminals are independent of each other and do not affect each other. Thus, the resource orchestration problem for GainNet can be modeled as a fully cooperative multi-agent Markov decision process (MDP). Therefore, we propose a DRL-based resource orchestration method to automatically learn the matching trajectory that maximizes the system GAI gain to achieve the optimal graph matching between the resources of the 6G terminals and the various-domain edge models, to obtain the optimal solution of GAI-oriented resource orchestration. The soft actor-critic (SAC) algorithm is selected as the approach for resource orchestration due to its high stability and high sample utilization, and the state space, action space, and reward are defined as follows.

$\bullet$ \textbf{State space}. The state space consists of two parts, including (a) the set of vertex features of all 6G end devices and (b) the feature vectors of all potential edges in the current state.

$\bullet$ \textbf{Action space}. The action space of GainNet's resource orchestration is a set of integers representing the indexes of the domain-specific models that terminals choose to serve.

$\bullet$ \textbf{Reward}. The reward is defined as the sum of all the 6G end devices’ GAI gain.

Based on the proposed DRL-based resource orchestration, the resource orchestration of the current GR is performed with the criterion of maximizing the long-term cumulative benefit, which can automatically learn the nonlinear mapping between the resource characteristics of the 6G terminals and the GAI edge models.

\section{Case Study}

\subsection{Effectiveness of GainNet}

\begin{figure*}[t]
\centering
\includegraphics[width=6.0in]{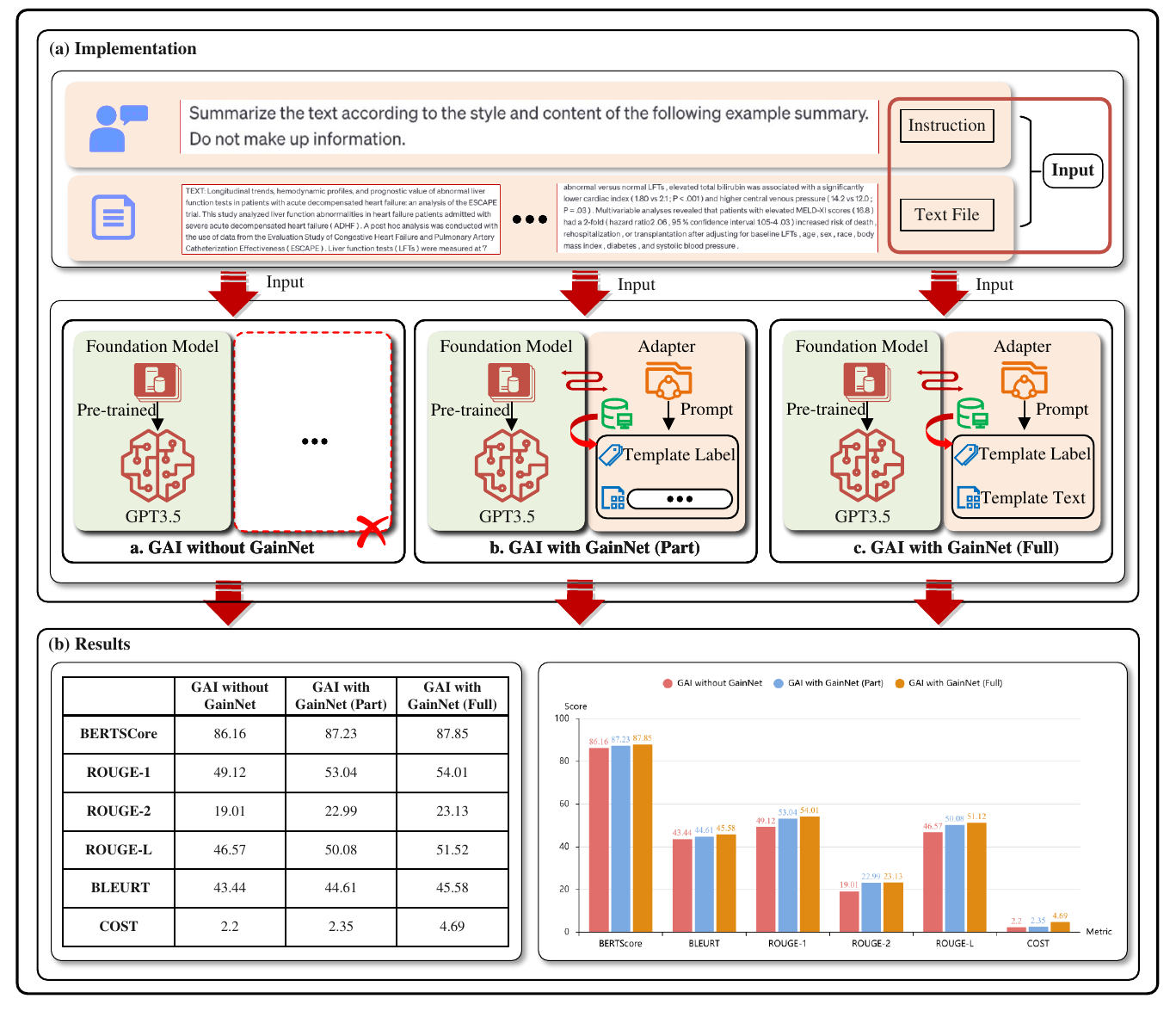}
\caption{Implementation and results of the case study in the medical domain. In the summary generation task, the user’s input consists of the instruction that expresses the intention and the text file, which is GAI’s executing object. Three methods in the implementation process correspond to the traditional method \textbf{a} which only uses the cloud foundation model (i.e., \textit{GAI without GainNet)}, the method \textbf{b}, where the edge-end model only gives partial prompt (template labels) (i.e., \textit{GAI with GainNet (Part)}), and the method \textbf{c}, where the edge-end model offers the full prompt (including the template labels and corresponding text) (i.e., \textit{GAI with GainNet (Full)}).}
\label{fig_4}
\end{figure*}

To demonstrate the effectiveness of the proposed GainNet framework, we take the summary generation task in the medical domain, which belongs to the field of natural language generation (NLG), as a case study. We selected the modified \textit{PubMed 20k RCT} as the data set \cite{ref7}. This dataset contains abstracts from approximately 20,000 medical randomized controlled trials and is widely regarded as the most authoritative source of evidence in the field of medicine. After removing some incomplete summaries, we have a training set (17,195), a validation set (300), and a test set (2,500). The sentences labeled background, goal, methods, and results in each abstract are combined into a single paragraph as the input text, while the sentences labeled conclusion are used as a reference summary of the medical abstract. We employ BERTScore, ROUGE series, and BLEURT to evaluate the generated summaries, where ROUGE mainly focuses on the lexical level overlap, while BERTScore and BLEURT leverage the BERT’s encoder to capture contextual information for a more in-depth evaluation at the semantic level. As can be seen from Fig. \ref{fig_4}, although the pre-trained GPT-3.5 with a large number of parameters already performs very well in text summarization, after using the edge-end model of GainNet to supplement the knowledge acquired from prompt learning in local data, all parameters are significantly improved. It is worth noting that when the prompt content of the edge-end model contains the template text, the cost of using the cloud GAI foundation model will increase significantly because it imposes additional requirements on the GAI model. Therefore, a good trade-off between performance and cost needs to be considered.

\subsection{Effectiveness and Robustness of GaiRom-ISCC}

\begin{figure*}[t]
\centering
\includegraphics[width=6.0in]{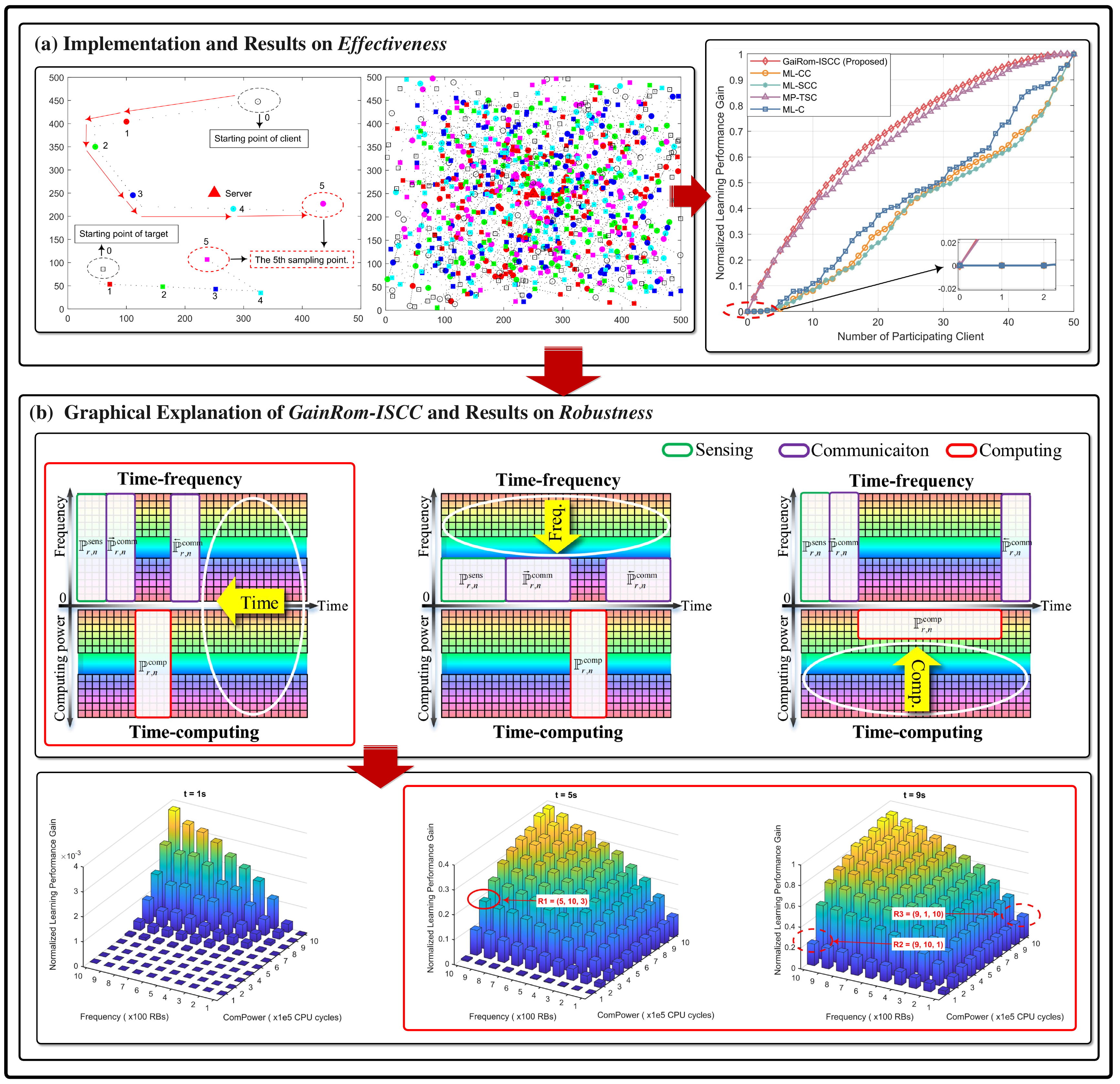}
\caption{Implementation and results of the case study in the traffic domain. Consider a square area of $500m\times 500m$, where connected intelligent vehicles act as clients and other vehicles act as the targets. The client performs FL-based distributed model training based on the target-related sensing data samples. CR is taken as the period for sampling, and Fig \ref{fig_5} (a) shows the sampling results of 50 vehicles and 100 targets in 5 consecutive CRs represented by specific colors and demonstrates the effectiveness of the proposed GaiRom-ISCC. Furthermore, Fig. \ref{fig_5} (b) gives the graphical explanation of the proposed GaiRom-ISCC, and proves the robustness of the resource orchestration based on GaiRom-ISCC.}
\label{fig_5}
\end{figure*}

Furthermore, to prove the effectiveness and robustness of the proposed GaiRom-ISCC in the resource provisioning of GainNet, as shown in Fig. \ref{fig_5}, we study the use case of intelligent vehicle network, a representative application scenario of edge AI in the traffic domain, where edge-end models are trained using the FL framework \cite{ref16}. It can be seen from Fig. \ref{fig_5} that, compared to resource orchestration mechanisms with 1) minimum latency in communication (ML-C), 2) minimum sum of latency in communication and computing (ML-CC), 3) minimum sum of latency in sensing, communication, and computing (ML-SCC), and 4) maximum product of the number of sensed targets and the sensing capacity (MP-TSC), the proposed GaiRom-ISCC can effectively improve the QoS of GAI. In addition, as shown in Fig. \ref{fig_5}, in the scenario with strict delay constraint (i.e., limited time resource), the resource orchestration strategy is changed, in which the time resource consumption is reduced from 9 to 5, but the same QoS of GAI can still be obtained as before the change of resource orchestration, which is similar to that in the scenario with limited frequency resources and computing power resources. Therefore, GaiRom-ISCC can adjust the resource orchestration strategies flexibly to guarantee the QoS of GAI robustly in the GainNet in the face of the end devices’ limited and highly dynamic resources.

\section{Challenges and Future Directions}

\subsection{Privacy Concerns and Social Issues}

As GAI continues to penetrate various fields, it gradually raises concerns about privacy concerns and social issues. First, attention should be given to privacy leakage and data security during the processes of pre-training, fine-tuning, and inference. Second, there are growing social concerns related to issues such as prejudice, ethics, and intellectual property infringement. Thus, it is imperative to study privacy protection strategies for GAI, such as differential privacy, while enhancing legislative supervision and developing technologies that can enhance the traceability of GAI models, such as blockchain-based technologies \cite{ref1}.

%

\subsection{Incentive Mechanism Design}

Effective and fair incentive mechanisms are essential to encourage the wide participation of 6G end devices in the optimization of the GAI model. Comprehensive consideration of resource contribution and data differences ensures that the value created by each participant matches the return obtained \cite{ref2,ref8}. The mathematical tools of economics that can be utilized include game theory, auction theory, contract theory, and other theories \cite{ref5}.

\subsection{Joint Optimization and Application}

The lifecycle of GAI is composed of optimization (i.e. pre-training and fine-tuning) and application (i.e. inference). If the optimization of the model is guided by the application goal, the results will be closer to the application requirements. Therefore, in addition to the resource management with ISCC proposed in this paper during the optimization stage, it is also necessary to coordinate and manage both the optimization and application stages of the GAI model \cite{ref10}.

\section{Conclusion}

In this paper, we propose the GainNet framework for mutualism between GAI and 6G networks, which realizes positive closed-loop knowledge flow and sustainable-evolution GAI model optimization based on the collaboration of cloud-edge-end intelligence. On this basis, a generic resource orchestration mechanism that integrates sensing, communication, and computing is proposed, where the nonlinear mapping is established between the local resources of 6G end devices and GAI models of various domains, which kills two birds with one stone, i.e., user association and resource allocation are implemented simultaneously. We also demonstrate the effectiveness of the proposed schemes by two simple case studies. Finally, we discuss the key challenges and future directions of the interplay between GAI models and 6G networks.

\vfill

\end{document}